\pdfoutput=1

\documentclass[aps,prb,reprint,showpacs,superscriptaddress]{revtex4-1}
\usepackage{graphicx}
\usepackage{amsmath}
\usepackage{amssymb}
\usepackage{amsfonts}
\usepackage{dcolumn}
\usepackage{dsfont}
\usepackage{latexsym}
\usepackage{rotating}
\usepackage{color}
\usepackage{latexsym}
\usepackage{bbm}
\usepackage{subfigure}
\usepackage{float}
\usepackage{epsfig}
\usepackage{psfrag}
\usepackage{natbib}
\usepackage{bm}
\usepackage{amsthm}
\usepackage{eucal}
\usepackage{mathrsfs}
\usepackage{url}
\usepackage{amssymb}
\usepackage{lipsum} % just for the example

\usepackage{color} % use colored text in latex

\usepackage{hyperref}
\hypersetup{
colorlinks=true,final=true,
        linkcolor=blue,
        citecolor=blue,
        filecolor=blue,
        urlcolor=blue,
}
\begin{document}

\title{Anomalous Nernst effect in type-II Weyl semimetals}

\author{Subhodip Saha}
\affiliation{Department of Physics, Indian Institute of Technology, Kharagpur,721302, India}
\author{Sumanta Tewari}
\affiliation{Department of Physics, Indian Institute of Technology, Kharagpur,721302, India}
\affiliation{Department of Physics and Astronomy, Clemson University, Clemson, South Carolina 29634, US}

%\date{\today}

%\pacs{31.15.A-, 71.20.Gj, 72.15.Gd, 03.65.Vf}

\begin{abstract}
{
 Topological Weyl semimetals (WSM), a new state of quantum matter with gapless nodal bulk spectrum and open Fermi arc surface states, have recently sparked enormous interest in condensed matter physics.
 Based on the symmetry and fermiology, it has been proposed that WSMs can be broadly classified into two types, type-I and type-II Weyl semimetals. While the undoped, conventional, type-I WSMs have point like Fermi surface and vanishing density of states (DOS) at the Fermi energy, the type-II Weyl semimetals break Lorentz symmetry explicitly and have tilted conical spectra with electron and hole pockets producing finite DOS at the Fermi level.
 The tilted conical spectrum and finite DOS at Fermi level in type-II WSMs have recently been shown to produce interesting effects such as a chiral anomaly induced longitudinal magnetoresistance that is strongly anisotropic in direction and a novel anomalous Hall effect.
 In this work, we consider the anomalous Nernst effect in type-II WSMs in the absence of an external magnetic field using the framework of semi-classical Boltzmann theory. Based on both a linearized model of time-reversal breaking WSM with a higher energy cut-off and a more realistic lattice model, we show that the anomalous Nernst response in these systems is strongly anisotropic in space, and can serve as a reliable signature of type-II Weyl semimetals in a host of magnetic systems with spontaneously broken time reversal symmetry.
 }
\end{abstract}
%\pacs{73.20.-r,  73.50.-h,  71.18.+y, 31.15.V-}

\maketitle

\section{Introduction}

The theoretical prediction and the subsequent experimental discovery of topological insulators \citep{wsm1,wsm45,wsm46,wsm47,wsm48,wsm49}revolutionized the band theory of solids. Topological insulators exhibit a full spectral gap in the bulk but robust gapless conducting states with non-degenerate spin texture protected by topology on the surface. Such topological systems with full spectral gap in the bulk have been classified into ten symmetry classes based on the presence or absence of discrete non-spatial symmetries such as time reversal, particle-hole, and the chiral symmetry \citep{wsm39, wsm40, wsm41, wsm42}. It has also been shown recently \citep{wsm3,wsm4,wsm5,wsm6,wsm7,wsm8,xu2015,wsm9,wsm10,wsm11,wsm12,wsm13,wsm14} that topological states of matter can exist even when the bulk spectrum is gapless, and these systems fall outside the ten-fold classification of topological insulators and superconductors. Three dimensional topological Weyl semimetals (WSM), a new state of quantum matter with gapless spectrum at bulk nodal points and spin non-degenerate open Fermi arc states on the surface, fall into this class and have recently sparked enormous interest in condensed matter physics \citep{wsm3,wsm4,wsm5,wsm6,wsm7,wsm8,xu2015,wsm9,wsm10,wsm11,wsm12,wsm13,wsm14}.

In WSMs a pair of non-degenerate energy bands touch at isolated points in the momentum space, and the low energy excitations near the band touching points, known as Weyl points, disperse linearly along all three momentum directions. The Weyl points act as source and sink of Abelian Berry curvature, which is an analog of magnetic field but defined in the momentum space \citep{wsm21,wsm22,wsm23,wsm24}. Weyl semimetals are different from Dirac semimetals, which are topologically protected in the presence of time reversal, space inversion and a lattice symmetry of the underlying crystal, in that WSMs violate space inversion and/or time reversal symmetry and are topologically protected by a non-zero quantized flux of Berry curvature across the Fermi surface known as the Chern number \citep{wsm34,volovik}. The Chern number is equal to the strength of the Berry magnetic monopole enclosed by the Fermi surface, while the  monopole charge (also called chirality) summed over all the Weyl points in the Brillouin zone vanishes \citep{wsm32,wsm33}. In recent studies several materials have been theoretically predicted to be Weyl semimetals, with experimental confirmation in several candidate materials such as, TaAs, NbAs, TaP, NbP \citep{wsm5,wsm6,wsm7,wsm8,wsm28,wsm29,wsm30,wsm31}.

In WSMs violation of Lorentz invariance, existence of non-trivial Berry curvature in the momentum space, and the (approximate) separate conservation of the number of Weyl electrons of different chiralities, lead to many anomalous transport and optical properties, such as anomalous Hall effect, dynamic chiral magnetic effect which is related to optical gyrotropy, and most importantly, nagative longitudional magnetoresistance in the presence of parallel electric and magnetic fields, due to non-conservation of separate electron numbers of opposite chirality for relativistic massless Fermions, an effect known as the chiral Adler-Bell-Jackiw anomaly \citep{volovik,wsm3,xu2011,wsm32,wsm33,bell1969,xu2011,wsm13,goswami2015,aji2012,adler1969}.

In recent work \citep{xu2015,wsm9} it has been proposed that based on the symmetry and fermiology WSMs can be broadly classified into two types, type-I and type-II Weyl semimetals. While the conventional type-I WSMs have point-like Fermi surface and vanishing density of states at the Fermi energy, the WSMs of type-II break Lorentz symmetry explicitly, resulting in a tilted conical spectra with electron and hole pockets producing finite density of states at the Fermi level \citep{xu2015,wsm9,wsm44}. The tilting can be generated in many different ways, e.g., by change in chemical doping or strain in different directions \citep{wsm37}. The tilted conical spectra and the finite density of states at the Fermi level in type-II WSMs have been shown to produce interesting effects such as chiral anomaly induced longitudinal magnetoresistance which is strongly anisotropic in space and a novel anomalous Hall effect\citep{wsm16,wsm17}. In this work we consider the anomalous Nernst effect in type-II Weyl semimetals in the framework of semi-classical Boltzmann theory\citep{wsm16,wsm19,wsm15,lundgren2014,kim2014,son2013}. Based on both a linearized dispersion model of time-reversal breaking WSMs and the corresponding lattice model, we show that the anomalous Nernst response in type-II WSMs is finite, sharply increases with decreasing chemical potential, and is strongly anisotropic in space, which can serve as a reliable signature of type-II Weyl semimetals in a host of magnetic systems with spontaneously broken time reversal symmetry.

The rest of the paper is organized as follows: In section II we introduce a linearized dispersion model for type-II WSMs with a higher energy cut-off. In section III we briefly discuss the derivation of the semiclassical Boltzmann distribution formula which we have used for calculating the anomalous Nernst response. The results of our calculation of Nernst signal and anisotropy have been included in section IV, where we have considered both the low energy linearized Hamiltonian of a type-II WSM as well as a more realistic lattice model. We include discussion of our central results in section IV. We end with a brief conclusion in section V.

\section{Low energy linearized Hamiltonian for topological WSM}
The low-energy linearized model for a time reversal broken type-II WSM with a single pair of Weyl points separated and tilted along the $k_z$ direction is described by the Hamiltonian,
\begin{equation}
H_{1,2}(\bm{k})=\hslash C_{1,2}(k_{z}\mp Q)\mp \hslash v \bm{\sigma}.(\bm{k}\mp Q\bm{e_{z}})
\label{Eq:H1}
\end{equation}
%%%%%%%%%%%%%%%%%%%%%%%%%%%%%%%%%%%%%%%%%%%%%%
\begin{figure}[h]
\begin{frame}{}
\hbox{\hspace{-3ex}\includegraphics[width=.52\textwidth]{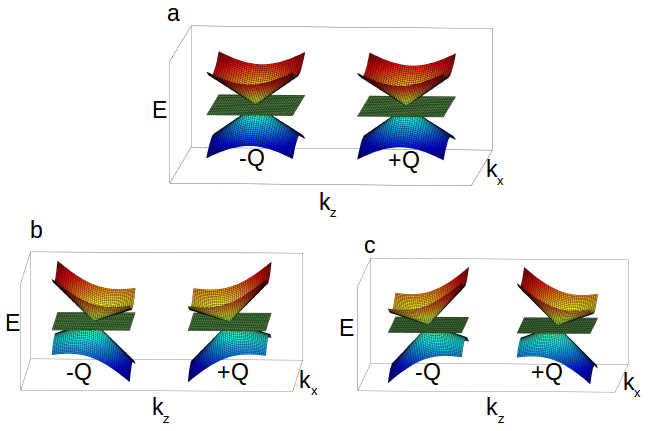}}
\end{frame}
\caption{Schematic illustrations for type-I WSM (a) Two untilted Weyl cones of type-I with the Weyl points located at $+Q$ and $-Q$ along the $k_z$ axis and tilting parameters $C_1=C_2=0$. (b) Weyl cones of type-I located at $\pm Q$ along the $k_z$ axis with tilting parameters $C_1/v=0.5, C_2/v=-0.5$. (c) Weyl cones of type-I located as before with tilting parameters $C_1/v=-0.5, C_2/v=0.5$}
\label{Fig:TypeI}
\end{figure}
%%%%%%%%%%%%%%%%%%%%%%%%%%%%%%%%%%%%%%%%%%%%%%%%%%%
\begin{figure}[h]
\begin{frame}{}
\hbox{\hspace{-1ex}\includegraphics[width=.52\textwidth]{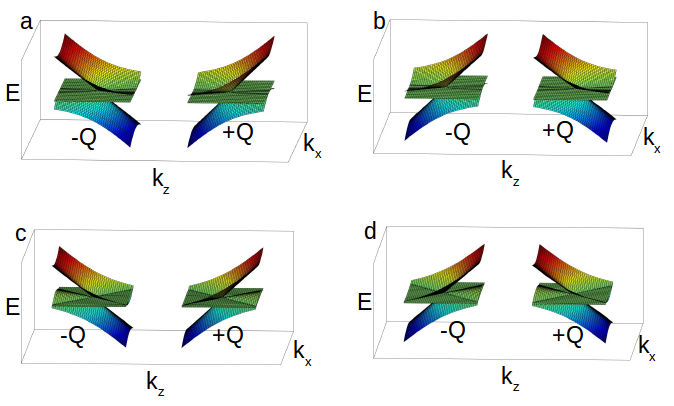}}
\end{frame}
\caption{Schematic illustrations for type-II Weyl cones. (a) and (b) are at the critical point between transition from type-I to type-II WSM with Weyl cones just touching the Fermi surface. Tilting parameters are taken as $C_1/v=1$ and $C_2/v=-1$ (a) and $C_1/v=-1, C_2/v=1$ (b). (c) and (d) are WSMs of type II. Weyl cones have gone through the Fermi surface and there is finite density of states at the Fermi level arising from electron and hole pockets coexisting with Weyl points. Tilting parameters are taken $C_1/v=1.3, C_2/v=-1.3$ (c) and $C_1/v=-1.3, C_2/v=1.3$ (d)}
\label{Fig:TypeII}
\end{figure}
where $\bm\sigma$'s are the Pauli matrices and $\sigma_0$ is the $2\times 2$ unit matrix. As unbounded linear dispersion model is not realistic we have used higher energy and momentum cut-offs in the following calculations. Fermi velocity of electrons in the absence of tilting is written as $v$. For numerical calculations the Fermi velocity $v$ is chosen as $10^6 $ m/s and the energy cutoff taken as $0.3$ eV. In Eq.~(\ref{Eq:H1}) the two Weyl points of opposite chirality are located at $(0,0,\pm Q)$ in the momentum space. We assume the momentum space splitting between the Weyl points as $Q=5 $ nm. The type of Weyl points is defined by the tilting parameters $C_{1,2}$. For $|C_{1,2}| < v$ the Weyl cones do not touch the Fermi surface. So for these values of the tilting parameters the system is a WSM of type-I. When $|C_{1,2}| = v$, the Weyl cones touch Fermi surface and is the critical point between transition from WSM-I to WSM-II. Finally, for $|C_{1,2}| > v$ the Weyl cones go inside the Fermi surface and the system is a WSM of type-II with electron and hole pockets and a finite density of states at the Fermi level. In Fig.~(\ref{Fig:TypeI} b,c) we have shown WSM of type-I where the Weyl cones, though tilted, do not touch the Fermi surface. In Fig.~(\ref{Fig:TypeII} a,b) Weyl cones touch the Fermi surface so are at the critical point between transition from type-I to type-II. Fig.~(\ref{Fig:TypeII} c,d) show type-II WSM with electron and hole pockets coexisting with topologically protected Weyl points. In the following, for calculations using the linearized dispersion model we will take $C_2 = -C_1$, as is usually the case (but not always, see Sec.~IV C) when the linearized model arises from a microscopic lattice model of type-II WSMs.

\section{Nernst effect in the presence of Berry curvature}
Generally Nernst response can be seen as a generation of electric field in the presence of a transverse temperature gradient. Conventionally Nernst effect can be observed in the presence of an external magnetic field transverse to the applied temperature gradient, but a non-trivial Berry curvature \citep{wsm21,wsm22,wsm23,wsm24,wsm26,wsm36} can also give rise to anomalous contribution to Hall and the Nernst signal. In the presence of electric field ($\textbf{E}$) and temperature gradient ($\bm\nabla$T) the linear response equations can be written as,
\begin{equation}
%\[
%M=
\begin{bmatrix}
\textbf{J} \\ \textbf{Q}
\end{bmatrix}
=
  \begin{bmatrix}
    \bm\sigma & \bm\alpha \\
    \bm{\bar{\alpha}} & \bm\kappa
  \end{bmatrix}
  \begin{bmatrix}
  \bm{E}\\
  -\bm{\nabla} T
  \end{bmatrix}
%\]
\label{Eq:Nernst}
\end{equation}

From Eq.~(\ref{Eq:Nernst}) the Nernst coefficient can be defined as,
\begin{equation}
\nu= \frac{E_y}{(-\frac{dT}{dx})}=\frac{\alpha_{xy} \sigma_{xx}-\alpha_{xx} \sigma_{xy}}{\sigma_{xx}^2+\sigma_{xy}^2}\\
\end{equation}
where $(\frac{dT}{dx})$ is the applied temperature gradient in $\hat{x}$ direction and $E_y$ is the electric field generated in the $\hat{y}$ direction. Usually, the longitudinal conductivity $\sigma_{xx}$ is much greater than the Hall conductivity $\sigma_{xy}$ ($\sigma_{xx}>>\sigma_{xy}$), as is the case in this paper, so we neglect $\sigma_{xy}^2$ in the denominator.
With this approximation, the expression for the Nernst coefficient becomes \\
\begin{equation}
\nu=\frac{\alpha_{xx}}{\sigma_{xx}}\left(\theta_P-\theta_H\right)\\
\label{Eq:Theta}
\end{equation}
where $\theta_H=\sigma_{xy}/\sigma_{xx}$ is called the Hall angle and $\theta_P=\alpha_{xy}/\alpha_{xx}$ is called the Peltier angle.
%As, in majority of cases $\theta_H $ is very small w.r.t $\theta_H$ so in many literature $\theta_H$ has been neglected.

It is now well known that the presence of non-trivial Berry curvature in systems with broken time reversal symmetry can give rise to an anomalous contribution to the Hall signal even in the absence of an external magnetic field. In systems with broken inversion symmetry but in which the time reversal symmetry remains unbroken the anomalous contribution to the Hall effect vanishes.
%So the system shows both conventional as well as anomalous contribution.
In the presence of Berry curvature, the semi-classical equations of motion for the position coordinate of an electron wave packet is given by,\\
\begin{equation}
\bm{\dot{r}}=\frac{1}{\hslash}\frac{\partial{\varepsilon(\bm k)}}{\partial \bm k}+
\frac{\dot{\bm p}}{\hslash}\times \bm\Omega_k\\
\label{Eq:req}
\end{equation}
Similarly the equation of motion for the momentum coordinate in the presence of both electric and magnetic fields is given by,
\begin{equation}
\bm{\dot{p}}=e\bm{E}+e\bm{\dot{r}}\times \bm{B}
\label{Eq:peq}
\end{equation}
It is clear from Eqs.~(\ref{Eq:req},\ref{Eq:peq}) that a non-zero Berry curvature $\Omega_k$ acts as a pseudo magnetic field in the equations of motion of an electron, albeit one that is defined in the momentum space.

 Solving the coupled equations, Eqs.~(\ref{Eq:req},\ref{Eq:peq}), for $\bm{\dot{r}}$ and $\bm{\dot{p}}$ and simplifying them we get,\\
\begin{equation}
\bm{\dot{r}}=D(\bm B,\bm \Omega_k)\left(\bm v_k+\frac{e}{\hslash}(\bm E\times \bm\Omega_k)+\frac{e}{\hslash} (\bm v_k.\bm\Omega_k)\bm B\right)\\
\label{Eq:req1}
\end{equation}
\begin{equation}
\bm{\dot{p}}=D(\bm B,\bm\Omega_k)\left(e\bm E+\frac{e}{\hslash}(\bm v_k\times \bm B)+\frac{e^2}{\hslash}(\bm E.\bm B)\bm\Omega_k\right)\\
\label{Eq:peq1}
\end{equation}
where, 
\begin{equation}
D(\bm B,\bm\Omega_k) =(1+e(\bm B.\bm\Omega_k)/\hslash)^{-1}
\label{Eq:Deq}
\end{equation}
 is a nontrivial phase-space factor arising from non-zero Poisson brackets of the coordinates \citep{duval2006}. In Eqs.~(\ref{Eq:req1},\ref{Eq:peq1}) we have used $\bm v_k=\hslash^{-1}\dfrac{\partial \varepsilon_{\bm k}}{\partial \bm k}$ as band velocity.

We now use Eqs.~(\ref{Eq:req1},\ref{Eq:peq1}) for $\bm{\dot{r}}$ and $\bm{\dot{p}}$ to solve the semi-classical Boltzmann equations in the relaxation time approximation \citep{wsm25,wsm35}, resulting in,
\begin{equation}
(\bm{\dot{r}}.\bm{\nabla_r}+\bm{\dot{k}}.\bm{\nabla_k})\\f_{\bm k}=-\frac{f_{\bm k}-f_{eq}}{\tau}
\label{Eq:Boltzmann}
\end{equation}
where $\tau$ is the scattering time of electrons, which for simplicity we take as independent of momentum. In Eq.~(\ref{Eq:Boltzmann}) $f_{\bm k}$ is the Fermi-Dirac distribution function with perturbation and $f_{eq}$ is the equilibrium Fermi-Dirac distribution. The linear response relations between the charge and the applied fields can be stated as,  
\begin{equation}
J_a=\sigma_{ab} E_b+\alpha_{ab} (-\nabla_b T)
\label{Eq:Jeq}
\end{equation}

We will consider the case when $\bm E=0$ and derive the longitudinal and transverse conductivities. We consider particular arrangement relevant for experimental measurements i.e. $\bm \nabla T=\nabla_x T \hat{x}, \bm B=B \hat{z}$ and $\bm E=0$. After making substitutions for $\dot r $ and $\dot p$ from Eqs.~(\ref{Eq:req1},\ref{Eq:peq1}) in Boltzmann Eq.~(\ref{Eq:Boltzmann}) we get,

%%%%%%%%%%%%%%%%%%%%%%%%%%%%%%%%%%%%%
%\iffalse
%The charge current in the presence of an electric field and a temperature gradient is given by  [\onlinecite{xiao2006berry}],
%\begin{multline}
%\bm J=-e \int {\bm[dk]}\left(\bm {v_k}+\frac{e}{\hslash}\bm E \times \Omega_{\bm k}\right) f_{\bm k}\\
%+\frac{k_B e \bm{\triangledown T}}{\hslash}\times \int {\bm[dk]}\Omega_{\bm{k}}s_{\bm{k}}
%\end{multline}
%In this expression $s_{\bm k}=-f_{\bm k}\log{f_{\bm k}}-(1-f_{\bm k})\log(1-f_{\bm k})$ is the entropy density, $\bm{[dk]}=\dfrac{d^3\bm{k}}{(2\pi)^3}$ denotes integration over the 3D momentum space.
%\fi
%%%%%%%%%%%%%%%%%%%%%%%%%%%%%%%%%%%%%%
%\iffalse
%\begin{multline}
%\bm Q=\int {\bm[dk]}(\epsilon_{\bm k}-\mu )\bm {v_k} f_{\bm {k}}+\frac{e}{\beta \hslash}\int {\bm[dk]}(\bm E \times \Omega_{\bm k}) s_{\bm k}\\
%+\frac{k_B  \bm{\triangledown T}}{\hslash}\times \int {\bm[dk]} \Omega_{\bm k}\left(\frac{\pi^2}{3}f_{eq}+\beta^2(\epsilon_k-\mu)^2 f_{eq}\right)\\
%-\frac{k_B  \bm{\triangledown T}}{\hslash}\times \int {\bm[dk]} \Omega_{\bm k}\left(\ln(1+e^{-\beta(\epsilon_k-\mu)^2}+2 Li_2 (1-f_{eq})\right)
%\end{multline}
%where $Li_2(z)$ is the polylogarithmic function of order 2, which is generally defined as,
%\begin{center}
%$Li_s(z)=\displaystyle \sum_{k=1}^\infty \frac{z^k}{k^s}$
%\end{center} 
%\fi
%%%%%%%%%%%%%%%%%%%%%%%%%%%%%%%%%%%%%

\begin{multline}
v_x\tau \nabla_x T\frac{\varepsilon-\mu}{T}\left(-\frac{\partial f_{eq} }{\partial \varepsilon}\right)+\frac{eB}{\hslash}\left(-v_x\frac{\partial }{\partial k_y}+v_y \frac{\partial}{\partial k_x}\right) f_{\bm{k}}\\=-\frac{f_{\bm {k}}-f_{eq}}{D(B,\Omega_k)\tau}
\label{Eq:Boltzmann1}
\end{multline}

We use the following expression of $f_k$ with correction factor $\Lambda$ for finite magnetic field,
\begin{multline}
f_{\bm {k}}=f_{eq}-\left(D\tau v_x \nabla_x T \frac{\varepsilon-\mu }{T}-\bm v.\bm \Lambda\right)\left(-\frac{\partial f_{eq}}{\partial \varepsilon}\right)
\label{Eq:Boltzmann2}
\end{multline}
Using this expression [Eq.~(\ref{Eq:Boltzmann2})] of $f_k$ in Eq.~(\ref{Eq:Boltzmann1}) we find,
\begin{multline}
\frac{eB}{\hslash}\left(v_y \frac{\partial}{\partial k_x}-v_x\frac{\partial}{\partial k_y}\right)\left(-D\nabla_x T\frac{\varepsilon-\mu}{T}v_x \tau+\bm v.\bm \Lambda\right)\\=-\frac{\bm v. \bm \Lambda}{\tau}
\end{multline} 

We find that $\Lambda_z=0$ as this equation must be valid for all values of $\bm v$. The equation can be simplified as,
\begin{multline}
eB\nabla_x T \frac{\varepsilon -\mu}{T}D\tau \left(\frac{v_x}{m_{xy}}-\frac{v_y}{m_{xx}}\right)+eB\left(\frac{v_y \Lambda_x}{m_{xx}}-\frac{v_x \Lambda_y}{m_{yy}}\right)\\
=-v_x \Lambda_x \left(-\frac{eB}{m_{xy}}+\frac{1}{D\tau}\right)-v_y \Lambda_y \left(\frac{eB}{m_{xy}}+\frac{1}{D\tau}\right)
\end{multline}
We now introduce two complex variables $V=v_x+i v_y$ and $\Lambda=\Lambda_x-i \Lambda_y$ to solve this equation and can be rewritten in this way,
\begin{multline}
Re\left[eB\tau D \nabla_x T\frac{\varepsilon-\mu }{T}V\left(\frac{1}{m_{xy}}+\frac{i}{m_{xx}}\right)\right]\\
=Re\left[V \Lambda\left(\frac{ieB}{m_{xx}}-\frac{1}{D\tau}\right)+\frac{eBV\Lambda}{m_{xy}}\right]
\end{multline}
For convenience of notation we define, $\Lambda_i=\tau \nabla_x T\dfrac{\varepsilon -\mu }{T}c_i$. We now write lengthy expressions for $c_x, c_y$ as,
%%%%%%%%%%%%%%%%%%%%%%%%%%%%%%%%%%%%%%%%%%%%%%%
\begin{widetext}
%\subsection*{Expressions for $ \bm {c_x},  \bm {c_y}$}
\begin{equation}
\bm {c_x}=eB D(B,\Omega_k)\dfrac{\left(\dfrac{v_x}{m_{xy}}-\dfrac{v_y}{m_{xx}}\right)\left(\dfrac{eB v_y}{m_{xx}}+\dfrac{eBv_x}{m_{xy}}+\dfrac{v_x}{D\tau}\right)-\left(\dfrac{v_y}{m_{xy}}+\dfrac{v_x}{m_{xx}}\right)\left(\dfrac{eBv_x}{m_{xx}}-\dfrac{v_y}{D\tau}-\dfrac{eBv_y}{m_{xy}}\right)}{\left(\dfrac{eBv_x}{m_{xy}}\right)^2-\left(\dfrac{v_x}{D\tau}+\dfrac{eBv_y}{m_{xx}}\right)^2-\left(\dfrac{eBv_x}{m_{xx}}-\dfrac{v_y}{D\tau}\right)^2+\left(\dfrac{eBv_y}{m_{xy}}\right)^2}
\label{Eq:cx}
\end{equation}

\begin{equation}
\bm {c_y}=eBD(B,\Omega_k) \dfrac{\left(\dfrac{v_x}{m_{xy}}-\dfrac{v_y}{m_{xx}}\right)\left(\dfrac{eBv_y}{m_{xy}}+\dfrac{eBv_x}{m_{xx}}-\dfrac{v_y}{D\tau}\right)-\left(\dfrac{v_y}{m_{xy}}+\dfrac{v_x}{m_{xx}}\right)\left(\dfrac{eBv_x}{m_{xy}}-\dfrac{v_x}{D\tau}-\dfrac{eBv_y}{m_{xx}}\right)}{\left(\dfrac{eBv_x}{m_{xx}}-\dfrac{v_y}{D\tau}\right)^2-\left(\dfrac{eBv_y}{m_{xy}}\right)^2-\left(\dfrac{eBv_x}{m_{xy}}\right)^2+\left(\dfrac{v_x}{D\tau}+\dfrac{eBv_y}{m_{xx}}\right)^2}
\label{Eq:cy}
\end{equation}
\end{widetext}
%%%%%%%%%%%%%%%%%%%%%%%%%%%%%%%%%%%%%%
In the presence of $\bm B$ and $\Omega_k$, the expression for the charge current $\bm J$ also modified\citep{xiao2006berry} by the factor $D(B,\Omega_k)$ [Eq.~\ref{Eq:Deq}]. The modified $\bm J$  can be rewritten as,
\begin{equation}
\bm J=-e\int[d \bm{k}]D^{-1} \bm{\dot r}f+\frac{k_B e \nabla T}{\hslash}\times\int [d \bm k]\Omega_{\bm k} s_{\bm k} 
\label{Eq:feq1}
\end{equation} 

Using expressions for $c_x, c_y$ [Eqs .~(\ref{Eq:cx},\ref{Eq:cy})] in the previously mentioned expressions of $f_k$ [Eq .~(\ref{Eq:Boltzmann2})], $f_k$ can be written in terms of $c_x, c_y$ and D as,
\begin{equation}
f_{\bm k}=f_{eq}-\left[\tau \nabla_x T \dfrac{\varepsilon- \mu }{T}\left(\frac{\partial f_{eq}}{\partial \varepsilon}\right)\right][(c_x-D)v_x+c_y v_y]
\label{Eq:Jeq1}
\end{equation}
Now substituting Eq.~(\ref{Eq:Jeq1}) in Eq.~(\ref{Eq:feq1}) and comparing with the previous linear response relation of J [Eq.~(\ref{Eq:Jeq})], we finally find thermal conductivities as,
\begin{equation}
\alpha_{xx}=e\int \bm{[dk]} v_x^2 \tau \frac{(\varepsilon-\mu)}{T}\left(-\frac{\partial f_{eq}}{\partial \varepsilon}\right)(c_x-D)
\end{equation}

\begin{multline}
\alpha_{xy}=e\int \bm{[dk]}\frac{(\varepsilon-\mu)}{T}\tau\left(-\frac{\partial f_{eq}}{\partial \varepsilon}\right)\left(v_y^2 c_y+v_x v_y(c_x-D)\right)\\
+ \frac{k_B e}{\hslash}\int [dk]\Omega_z s_k
\end{multline}

In the above expression, $v_x=\hslash^{-1}\dfrac{\partial \varepsilon}{\partial k_x}, v_y=\hslash^{-1}\dfrac{\partial \varepsilon}{\partial k_y}$ are the band velocities of electrons and $\bm {[dk]} = \dfrac{d^3 \bm k}{(2 \pi)^3}$ denotes integration over the 3D momentum space. 

Similarly the electrical conductivity tensors  in the presence of electric field can be obtained as,
\begin{equation}
\sigma_{xx}=-e^2\int \bm{[dk]} v_{x}^2\tau \left(-\frac{\partial f_{eq}}{\partial \varepsilon}\right)(c_x-D)
\end{equation}

\begin{multline}
\sigma_{xy}=-e^2\int \bm{[dk]}\left(-\frac{\partial f_{eq}}{\partial\varepsilon}\right) \tau\left(v_y^2 c_y+v_x v_y (c_x-D)\right)\\
+\frac{e^2}{\hslash}\int [dk]\Omega_z f_k
\end{multline}

 For realistic values of the chemical potentials $c_x, c_y, D$ reduce to, $c_x-D\rightarrow -1$ and $c_y \rightarrow \omega \tau$ where $\omega$ is the cyclotron frequency, $\omega=\frac{eB}{m}$ (m is the effective mass of electrons). By realistic values of chemical potential we mean $\mu$ in range of 15 meV-100 meV \citep{zhang2015tantalum,mar1992metal,wsm9}. According to this approximation longitudinal conductivity tensors can be simplified to,
 
\begin{equation}
\sigma_{xx}=e^2\int \bm{[dk]} v_{x}^2\tau \left(-\frac{\partial f_{eq}}{\partial \varepsilon}\right)
\label{Eq:Sigmax}
\end{equation}
\begin{equation}
\alpha_{xx}=-\frac{e}{T}\int \bm{[dk]} v_x^2 \tau (\varepsilon-\mu)\left(-\frac{\partial f_{eq}}{\partial \varepsilon}\right)
\label{Eq:Alphax}
\end{equation}

Using full expressions of $c_x, c_y$ [Eqs.~(\ref{Eq:cx},\ref{Eq:cy})], D [Eq.~(\ref{Eq:Deq})] and keeping terms only upto linear order in B the transverse conductivity tensors can be obtained as,

\begin{multline}
\sigma_{xy}=-\frac{e^3 \tau^2 B}{\hslash^2}\int \bm{[dk]}\left(-\frac{\partial f_{eq}}{\partial\varepsilon}\right) \left(\frac{v_x^2 \partial^2\varepsilon}{\partial k_y^2}-\frac{v_x v_y \partial^2 \varepsilon}{\partial k_x \partial k_y }\right)\\
+\frac{e^2}{\hslash}\int \bm{[dk]}\Omega_z f_k
\label{Eq:Sigmay}
\end{multline}
\begin{multline}
\alpha_{xy}=\frac{e^2 \tau^2 B}{T \hslash^2}\int \bm{[dk]}(\varepsilon-\mu)\left(-\frac{\partial f_{eq}}{\partial \varepsilon}\right)\left(\frac{v_x^2\partial^2\varepsilon}{ \partial k^2_y}-\frac{v_x v_y \partial^2 \varepsilon}{\partial k_x \partial k_y}\right)\\
+\frac{k_B e}{\hslash}\int \bm{[dk]}\Omega_z s_k
\label{Eq:Alphay}
\end{multline}

%\textcolor{red}{Replacing 1/m by $\frac{\partial^2 \epsilon}{\partial k_y^2}$ and $\frac{\partial^2 \epsilon }{\partial k_x \partial k_y}$}.
As in this paper we are focusing only on anomalous contributions, in the absence of an external magnetic field, the transverse conductivity tensors simplifies to,

\begin{equation}
\sigma_{xy}=\frac{e^2}{\hslash}\int \bm{[dk]}\Omega_z f_k
\label{Eq:Sigmay1}
\end{equation}

\begin{equation}
\alpha_{xy}= \frac{k_B e}{\hslash}\int \bm{[dk]}\Omega_z s_k
\label{Eq:Alphay1}
\end{equation}

For the calculations of the Berry curvatures we use the expression \citep{wsm26},\\\\
\begin{equation}
\Omega^{n}_{a,b}=i\sum_{m\neq n} \frac{\langle n\vert \frac{\partial H}{\partial k_{a}}\vert m\rangle \langle m\vert \frac{\partial H}{\partial k_b}\vert n\rangle -a\leftrightarrow b}{(\varepsilon_n-\varepsilon_m )^2}
\label{Eq:BC}
\end{equation}
where $\varepsilon_n$ is the energy of the $n$-th band and the sum is over all eigenstates $\vert m\rangle$ of the Hamiltonian excluding the eigenstate $\vert n\rangle$.
 From the denominator of the expression it is clear that the Berry curvature peaks up near the band touching points which in the present work are precisely the Weyl points. So in our calculations with the linearized model the Berry curvature has peaks at $(0,0,\pm Q)$ in the momentum space, which give rise to the anomalous Hall and Nernst effects even in the absence of an external magnetic field.

\section{Results}
In this section we first discuss the anomalous Nernst signal from the low energy linearized model with a high energy cutoff followed by the results from a lattice model of WSMs where the tilt is in the $k_z$ direction. We also point out the role of the entropy density $s_k$ in the results for the Nernst signal in the linearized model for opposite tilts of the Weyl cones. We then consider a lattice model for type-II WSMs with the tilt in the $k_x$ direction and illustrate the main result of this work, a significant anisotropy in the anomalous Nernst signal depending on the direction of the temperature gradient, which can be used as a signature of type-II Weyl semimetals with broken time reversal symmetry.
%and anisotropy signature meets to a certain point near phase transition from WSM-I to WSM-II.

\subsection{Anomalous Nernst Signal in the low energy linearized model}
The anomalous Nernst response in the presence of Berry curvature can be calculated from the Eqs.~(\ref{Eq:Theta},\ref{Eq:Sigmax},\ref{Eq:Alphax},\ref{Eq:Sigmay1},\ref{Eq:Alphay1},\ref{Eq:BC}).

\begin{figure}[h]
\begin{frame}{}
\hbox{\hspace{-3ex}\includegraphics[width=.53\textwidth]{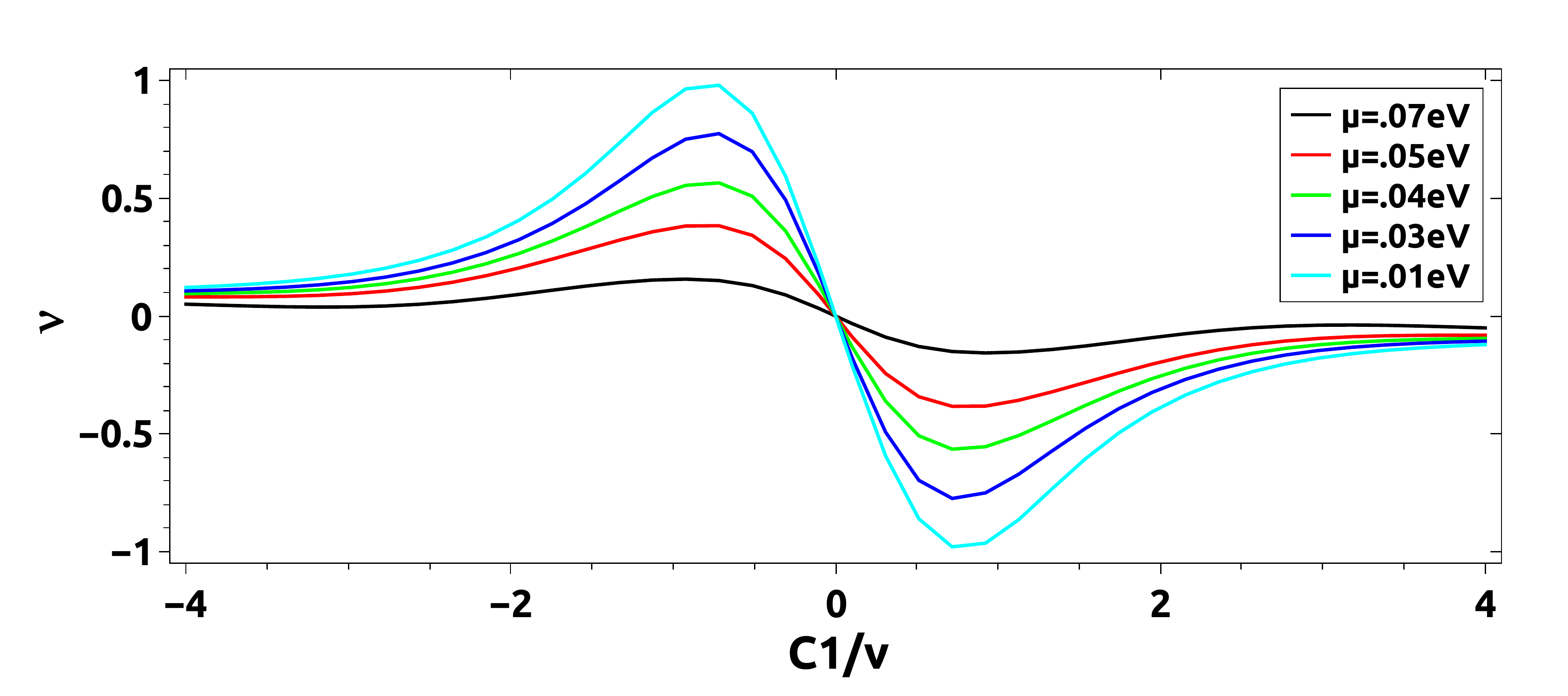}}
\end{frame}
\caption{Normalized anomalous Nernst coefficient ($\nu$) for a linearized model of WSM, Eq.~(\ref{Eq:H1}), as a function of the tilting parameter of Weyl cones. The Nernst signal has been calculated for different values of the chemical potential. When the tilting is zero the anomalous Nernst signal vanishes, which is an artifact of the linearized model (see Fig.~(\ref{Fig:NLattice})). The slopes of the curves change when the system makes the transition from type-I to type-II WSM, i.e., upto $|C_1|<v$ the slope is negative whereas for $|C_1|>v $  the slope is positive, which indicates a qualitative change in the behavior of the anomalous Nernst signal at the critical point $|C_1|=v$. It is important to note that for a fixed value of the tilting parameter the anomalous Nernst signal increases with decreasing values of the chemical potential, i.e., as the system approaches  the undoped limit}
\label{Fig:Nlinear}
\end{figure}

The anomalous Nernst signal ($\nu$) on the tilting parameter $C_1$ (we take $C_2 = -C_1$ in the linearized model) for a set of finite values of the chemical potential is shown in Fig.~(\ref{Fig:Nlinear}). As mentioned before, here we have introduced an energy and momentum cut-off as unbounded linear dispersion is not realistic for physical Weyl semimetals. In the realistic systems there will be intrinsic energy-momentum cutoff arising from the lattice parameter in a realistic lattice model description of the WSMs. Here in the approximate linearized model we observe that the anomalous Nernst signal has a negative slope with the tilting parameter for WSM type-I region from $C_1/v=-1$ to $C_1/v=1$. For WSM of type-II $C_1/v>1$ and $C_1/v<-1$ the Nernst signal has a positive slope with the tilting parameter. It is important to note that for a fixed value of the tilting parameter the anomalous Nernst signal increases with decreasing values of the chemical potential, i.e., with the system approaching the undoped limit. In the limit of $C_1/v\longrightarrow \infty $ or $C_1/v\longrightarrow -\infty$ the Nernst signal converges to a finite value which approximately vanishes.

%%%%%%%%%%%%%%%%%%%%%%%%%%%%%%%%%%%%%%%%%%%%%%%%%
\begin{figure}[h]
\begin{frame}{}
\hbox{\hspace{-3ex}\includegraphics[width=.5\textwidth]{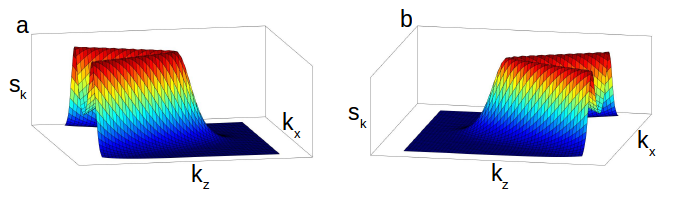}}
\end{frame}
\caption{Entropy density in the linearized model of a WSM. (a) shows the entropy density for the tilt parameter $C_1/v=-1.5$, and (b) shows the entropy density for the tilt parameter $C_1/v=1.5$. The entropy density moves significantly in the momentum space with change in the tilting of the Weyl cones from $C_1/v =-1.5$ to $C_1/v = 1.5$ and, coupled with the fact that the Berry curvature distribution is independent of the tilting parameter and depends only on the location of the Weyl nodes, results in opposite anomalous Nernst signal for opposite tilting of the Weyl cones. As it is a linearized model with a higher energy cut-off the plot of entropy density does not look like a periodic function.}
\label{Fig:Entropy}
\end{figure}
%%%%%%%%%%%%%%%%%%%%%%%%%%%%%%%%%%%%%%%%

The entropy density
\begin{equation}
 s_{\bm k}=-f_{\bm k} \log{f_{\bm k}}-(1-f_{\bm k})\log(1-f_{\bm k})
\end{equation}
 for the linearized model is shown in Fig.~(\ref{Fig:Entropy}). We observe a significant change in the entropy density along the $k_z$ axis. As in the low energy linearized model the Berry curvature does not change with the tilt of the Weyl cones, change in entropy density is what causes the opposite Nernst signal for $C_1/v=1.5$ and $C_1/v=-1.5$.
It can be seen from Fig.~(\ref{Fig:Nlinear}) that the anomalous Nernst signal vanishes \citep{wsm18} when there is no tilting. As has been discussed before \citep{wsm18} this is an artifact of the linearized model and a more realistic lattice model for the time reversal breaking WSM produces a non-zero anomalous Nernst signal even in the absence of tilting of the Weyl cones.
%   As a more realistic model than the linear cut off model we use a lattice model, we use a low energy TR broken lattice hamiltonian,

\subsection{Anomalous Nernst signal for a lattice model of time reversal breaking WSM}
For a lattice model of the time reversal breaking Weyl semimetal with an intrinsic energy-momentum cut-off provided by the lattice spacing we take the Hamiltonian,
\begin{multline}
H(\bm k)=\gamma (\cos(k_z)-\cos(k_0))\sigma_0\\
-\left( m[2-\cos(k_y)-\cos(k_x)]+2t_x[\cos(k_z)-\cos(k_0)]\right)\sigma_1\\
-2t\sin(k_y)\sigma_2+2t\sin(k_x)\sigma_3
\label{Eq:Hlattice}
\end{multline}
where $\gamma$ is the tilt parameter along the $k_z$ direction, $\sigma$'s are the Pauli matrices and $\sigma_0$ is the $2\times 2$ unit matrix. We take the lattice parameter  $k_0=\pi/2$, $t_x=t$ and $m=2t$ for our numerical calculations. The Hamiltonian in Eq.~(\ref{Eq:Hlattice}) produces two Weyl points located at $(0,0,\pm\pi/2)$ in the momentum space and makes a transition from type-I to type-II at $\gamma/t=\pm2$ when the chemical potential $\mu$ is zero.
%%%%%%%%%%%%%%%%%%%%%%%%%%%%%%%%%%%%%%%%%%%%%%
\begin{figure}[h]
\begin{frame}{}
\hbox{\hspace{-2ex}\includegraphics[width=.53\textwidth]{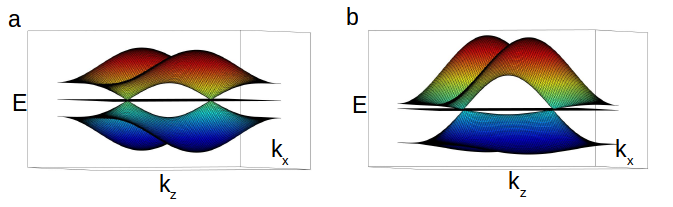}}
\hbox{\hspace{-4ex}\includegraphics[width=.53\textwidth]{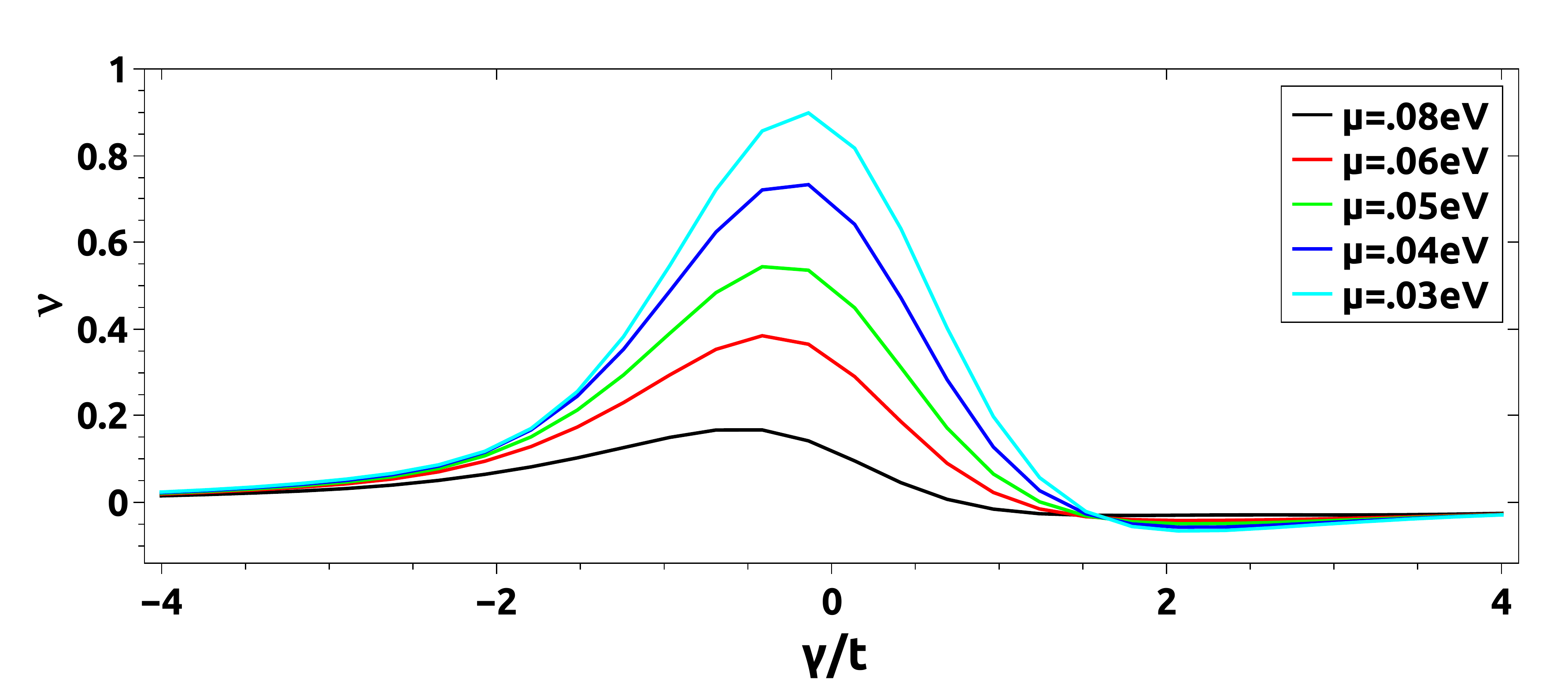}}
\end{frame}
\caption{Normalized anomalous Nernst signal for a lattice model of time reversal breaking WSM. Upper panel: (a) indicates the energy spectrum for a lattice model, Eq.~(\ref{Eq:Hlattice}), when there is no tilting in the Weyl cones ($\gamma=0$). (b) shows the Weyl cones tilted from each other in opposite direction with the tilting parameter $\gamma/t=1.5$. In this case the tilts are in $k_z$ direction similar to the linearized model, Eq.~(\ref{Eq:H1}). The Weyl points are located at $(0,0,\pm\pi/2)$ in the momentum space. Lower panel shows the anomalous Nernst coefficient($\nu$) for the lattice model as a function of the tilt parameter $\gamma$ of the Weyl cones. The Nernst signal has been calculated for different values of the chemical potential and increases monotonically with decrease in the chemical potential. The anomalous Nernst signal is finite even in the absence of tilting of the Weyl cones \citep{wsm18}, in contrast to the case of the linearized model (see Fig.~(\ref{Fig:Nlinear})). But near the end points i.e.  with very large tilting, the signals from both the linearized model and the lattice model matche with each other and the anomalous Nernst signal approximately vanishes.}
\label{Fig:NLattice}
\end{figure}
%%%%%%%%%%%%%%%%%%%%%%%%%%%%%%%%%%%%%%%%%%%%%%%%%%
We calculate Nernst signal for this lattice model according to Eqs~.(\ref{Eq:Theta},\ref{Eq:Sigmax},\ref{Eq:Alphax},\ref{Eq:Sigmay1},\ref{Eq:Alphay1},\ref{Eq:BC}). The dependence of the anomalous Nernst signal on the tilt parameter $\gamma$ has been plotted in Fig.~(\ref{Fig:NLattice}) for different values of the chemical potential. We find that the lattice model results in the anomalous Nernst effect which is considerably different from that in the linearized model. The most important difference is at zero tilting ($\gamma=0$), for which we find a finite Nernst signal (in fact a peak plotted as a function of $\gamma$), in contrast to zero Nernst signal for the untilted Weyl nodes in the linearized model. Fig.~\ref{Fig:NLattice}, therefore, establishes the fact that the vanishing of the anomalous Nernst effect for $\gamma=0$ is an artifact of the linearized model. This result agrees with Ref .~[\onlinecite{wsm18}], where only the case of $\gamma=0$ was considered, and disagrees with the results for anomalous Nernst signal in Ref .~[\onlinecite{lundgren2014}]. We also observe that for $\gamma/t\longrightarrow \infty$ or $\gamma/t\longrightarrow-\infty$ the Nernst signal converges to a finite value which approximately vanishes as in the linearized model. It is important to note that, as in the linearized model, for a fixed value of the tilting parameter the anomalous Nernst signal monotonically increases with decreasing values of the chemical potential.

\subsection{Anisotropy in the anomalous Nernst signal for type-II WSM}
As the previous models were symmetric with respect to $k_x, k_y$ there was no anisotropy in the anomalous Nernst signals between the cases when the temperature gradient is applied along $\hat{x}$ or $\hat{y}$ axes. To illustrate the anisotropy when the temperature gradient is applied along and perpendicular to the tilt axis,
we now consider a lattice model for a time reversal broken Weyl semimetal with the Weyl cones tilted along the $k_x$ axis,
\begin{multline}
H(\bm k)=\gamma (\sin(k_x)-\sin(k_0))\sigma_0\\
-\left( m[2-\cos(k_y)-\cos(k_x)]+2t_x[\cos(k_z)-\cos(k_0)]\right)\sigma_1\\
-2t\sin(k_y)\sigma_2+2t\sin(k_x)\sigma_3
\label{Eq:HLattice1}
\end{multline}
Here, as before, $\sigma$'s are conventional Pauli matrices, $\sigma_0$ is $2\times 2$ unit matrix, and $\gamma$ is the tilt parameter along the $k_x$ direction. We take the lattice parameters $k_0=0$, $t_x=t$, and $m=2t$. The Hamiltonian in Eq.~(\ref{Eq:HLattice1}) produces two Weyl cones at $(0,0,\pm \pi/2)$ still separated along the $k_z$ axis but now tilted along $k_x$.\\
%%%%%%%%%%%%%%%%%%%%%%%%%%%%%%%%%%%%%%%%
\begin{figure}[h]
\begin{frame}{}
\hbox{\hspace{-3ex}\includegraphics[width=.53\textwidth]{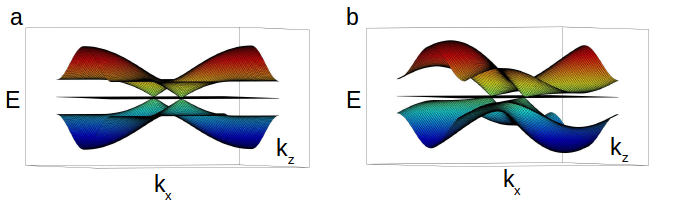}}
\hbox{\hspace{-5ex}\includegraphics[width=.53\textwidth]{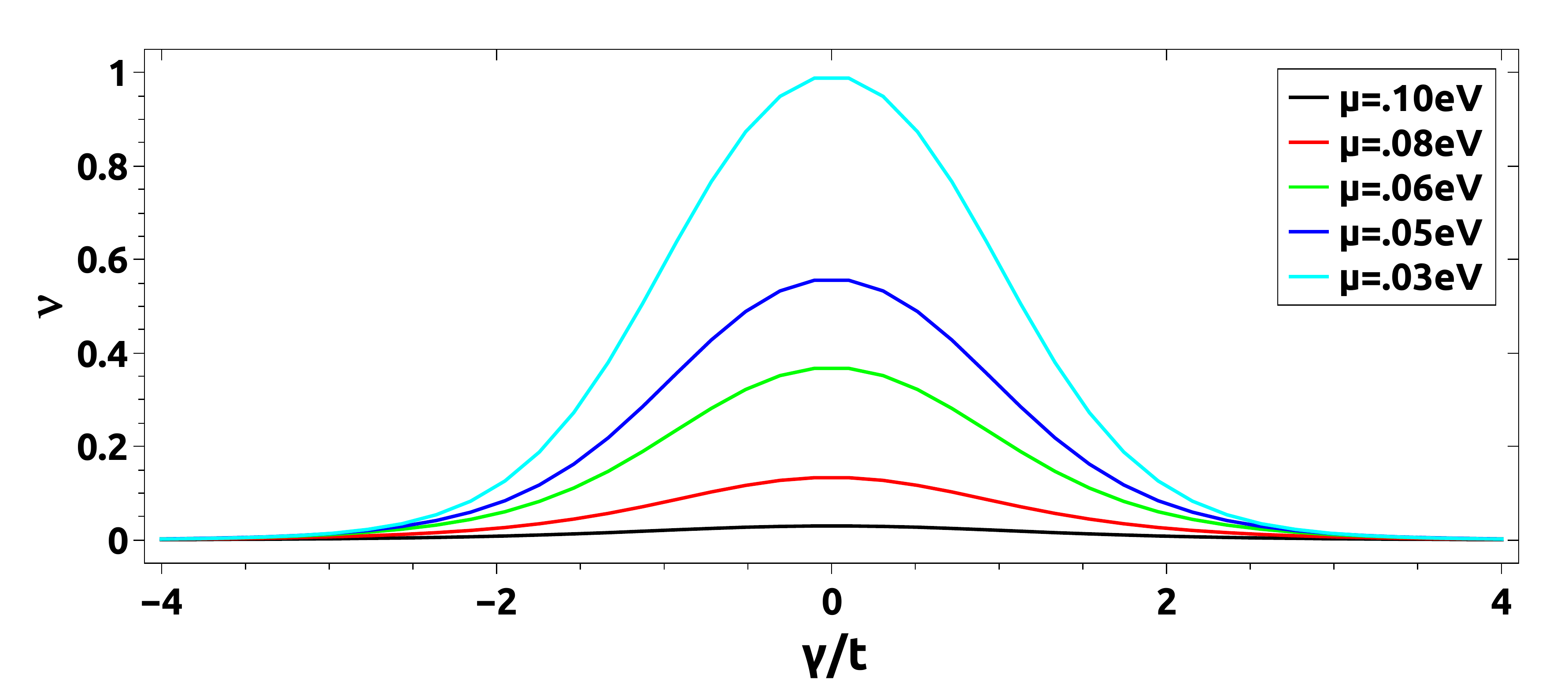}}
\end{frame}
\caption{Normalised anomalous Nernst signal for a lattice model of type-II WSM. Upper panel: (a) indicates the energy spectrum of lattice model Eq.~(\ref{Eq:HLattice1}) when there is no tilting in the Weyl cones. (b) shows both Weyl cones tilted in the same direction along the $k_x$ axis when the tilt parameter $\gamma/t=1.5$. Lower panel: Anomalous Nernst coefficient ($\nu$) is plotted with the tilt parameter. The Nernst signal has been calculated for different values of the chemical potential. When the tilt is zero we get a finite Nernst signal similar to the previous lattice model with the tilt in the $k_z$ direction. As both the Weyl cones are tilted in the same direction, in contrast to Fig.~(\ref{Fig:NLattice}), the curves are symmetric about the tilting axis and the anomalous Nernst signal has the same value whether the tilting parameter is positive or negative.}
\label{Fig:Nlattice1}
\end{figure}
%%%%%%%%%%%%%%%%%%%%%%%%%%%%%%%%%%%%%%%%%%%%%%%
\begin{figure}[h]
\begin{frame}{}
\hbox{\hspace{-4ex}\includegraphics[width=.53\textwidth]{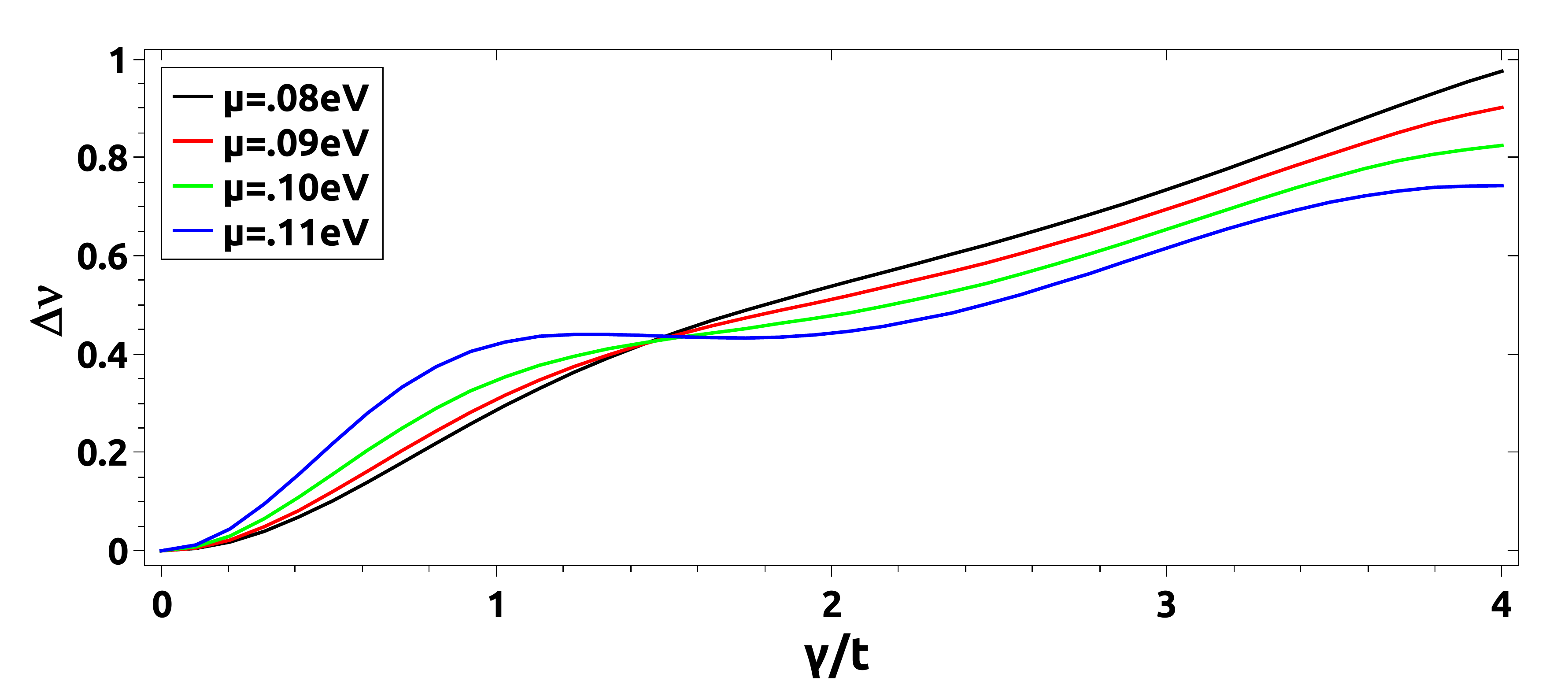}}
\end{frame}
\caption{Normalised difference in Nernst signal ($\Delta \nu$) between the cases when the temperature gradient is applied parallel and perpendicular to the tilt direction plotted as a function of the tilt parameter in time reversal breaking Weyl semimetal described in Eq.~(\ref{Eq:HLattice1}). The anisotropy has been calculated for different values of the chemical potential. As expected, when the tilt is zero there is no anisotropy and the anisotropy increases with increase in the tilt of the Weyl cones.}
\label{Fig:Anisotropy}
\end{figure}
%%%%%%%%%%%%%%%%%%%%%%%%%%%%%%%%%%%%%%%%%%%%%%%%%%%%%%%%%%%%%
The anomalous Nernst response in the presence of Berry curvature can be calculated from Eqs.~(\ref{Eq:Theta},\ref{Eq:Sigmax},\ref{Eq:Alphax},\ref{Eq:Sigmay1},\ref{Eq:Alphay1},\ref{Eq:BC}). The dependence of Nernst signal for Hamiltonian in Eq.~(\ref{Eq:HLattice1}) has been plotted in Fig.~(\ref{Fig:Nlattice1}). For  $\gamma/t\longrightarrow \infty$ or $\gamma/t\longrightarrow-\infty$ the anomalous Nernst signal converges to zero.  It is important to note that in the model of Eq.~(\ref{Eq:HLattice1}) both the Weyl cones are tilted in the same direction along the $k_x$ axis, which is different from the model in Eq.~(\ref{Eq:Hlattice}) where the tilts are in opposite directions. As the Weyl cones are tilted in the same direction, we get the same Nernst signal whether the tilt parameter is positive or negative, and in contrast to Fig.~(\ref{Fig:NLattice}) the curves in Fig.~(\ref{Fig:Nlattice1}) are symmetrical.

The Hamiltonian in Eq.~(\ref{Eq:HLattice1}) is not symmetric in $k_x$ and $k_y$, as the $\sigma_0$ term responsible for the tilting of the Weyl cones contains only $k_x$. Consequently, we expect the system to show strong anisotropy between the cases when the applied temperature gradient is parallel or perpendicular to the tilt axis.
For calculating the Nernst coefficient when the temperature gradient is along $x$ direction and the generated electric field is in the $y$ direction, we use the formula,
\begin{equation}
\nu_x= \frac{E_y}{-\frac{dT}{dx}}=\frac{\alpha_{xy} \sigma_{xx}-\alpha_{xx} \sigma_{xy}}{\sigma_{xx}^2+\sigma_{xy}^2}\\
\label{Eq:Theta1}
\end{equation}
Similarly, for calculating the Nernst signal with temperature gradient along $y$ direction and the generated electric field is in the $x$ direction, we use the expression,
\begin{equation}
\nu_y= \frac{E_x}{-\frac{dT}{dy}}=\frac{\alpha_{xy} \sigma_{yy}-\alpha_{yy} \sigma_{xy}}{\sigma_{yy}^2+\sigma_{xy}^2}\\
\label{Eq:Theta2}
\end{equation}
Here $\sigma_{xy}=\sigma_{yx}\quad $and$ \quad \alpha_{xy}=\alpha_{yx}$ due to symmetry in the anomalous terms.
Finally, we calculate $\Delta \nu= \nu_x-\nu_y$ as the signature of anisotropy in the anomalous Nernst signal in type-II Weyl semimetals that can be measured in experiments. In our model $\Delta \nu$ was calculated according to Eqs.~(\ref{Eq:Sigmax},\ref{Eq:Alphax},\ref{Eq:Sigmay1},\ref{Eq:Alphay1},\ref{Eq:BC},\ref{Eq:Theta1},\ref{Eq:Theta2}).
The anisotropy in the anomalous Nernst coefficient for this lattice model is illustrated in Fig.~(\ref{Fig:Anisotropy}). In the absence of tilting, i.e $\gamma/t=0$, we don't observe any anisotropy, as expected for type-I WSM without tilting of Weyl cones. We find that the anisotropy increases with the tilting of the Weyl cones as the direction of tilt introduces a preferred direction on the $x-y$ plane. Thus, the system responds differently when the applied temperature gradient is along the direction of tilt or perpendicular to it.
%Our result shows that near $\gamma/t=1.5$ which is the phase transition between type-I to type-II all the anisotropy signatures pass through the same point.

\section{Conclusion}

We study the anomalous Nernst signal of time reversal broken type-II Weyl semimetals for both low-energy linearized dispersion model with a high energy cut-off as well as for more realistic lattice models. We observe that the anomalous Nernst signal vanishes in the linearized model for zero tilting but for more realistic lattice models the anomalous Nernst effect is finite (in fact has a peak) when the tilt parameter vanishes. The vanishing of anomalous Nernst effect in the linearized model is thus an artifact of the low energy description and this agrees with a similar result in Ref.~[\onlinecite{wsm18}] (which treats only the case of zero tilting) and disagrees with Ref.~[\onlinecite{lundgren2014}]. We also study the anomalous Nernst signal in more realistic lattice models where the tilts are along the $k_z$ and $k_x$ axes. Generically, we find that the Nernst effect has a peak for zero tilting and decreases with increasing tilts of the Weyl cones. Importantly, we also find that the anomalous Nernst signal increases monotonically with decreasing values of the chemical potential, i.e., as the system approaches the undoped limit. A central result of this work is a pronounced anisotropy in the anomalous Nernst effect in type-II Weyl semimetals between the cases when the external temperature gradient is applied parallel and perpendicular to the tilt axis. The tilt in the Weyl cones introduces a preferred direction in momentum space and the corresponding symmetry breaking introduces strong anisotropy in topological response functions which can be used as an important marker for type-II WSMs in magnetic systems with spontaneously broken time reversal symmetry.  

\acknowledgements
The authors appreciate access to the computing facilities in the Department of Physics, IIT Kharagpur, India. We would like to thank G. Sharma, M. Chakraborty for some useful discussions.

\subsection*{Author contribution statement}
S Tewari conceived the problem, S Saha performed the calculations, S Saha and S Tewari wrote the paper.
\\\\

\textit{Note added:} After our paper was completed, we became aware of a recent preprint Ref.~[\onlinecite {ferreiros2017arxiv}], which also discusses anomalous Nernst effect in tilted Weyl semimetals, but only from a low energy linearised dispersion model.

\bibliography{wsmref}

\end{document}